\documentclass[%
 aip,
 amsmath,amssymb,
reprint,%
floatfix
]{revtex4-1}

\usepackage{graphicx}
\usepackage{dcolumn}
\usepackage{bm}

\usepackage[utf8]{inputenc}
\usepackage[T1]{fontenc}
\usepackage{mathptmx}
\usepackage[font=footnotesize]{caption}

\begin{document}

\preprint{}

\title[]{Control of magnetoelastic coupling in Ni/Fe multilayers using  He$^+$  ion irradiation}

\author{G. Masciocchi}
\email{gmascioc@uni-mainz.de}
 \affiliation{Institute of Physics, Johannes Gutenberg University
Mainz, Staudingerweg 7, 55099 Mainz, Germany}
 \affiliation{Sensitec GmbH, Walter-Hallstein-Straße 24, 55130 Mainz, Germany}
\author{J. W. van der Jagt}%
\affiliation{ 
Spin-Ion Technologies, 10 boulevard Thomas Gobert, 91120 Palaiseau, France 
}%
\affiliation{ Université Paris-Saclay, 3 rue Juliot Curie, 91190 Gif-sur-Yvette, France}%

\author{M.-A. Syskaki}
\affiliation{Institute of Physics, Johannes Gutenberg University
Mainz, Staudingerweg 7, 55099 Mainz, Germany}
\affiliation{Singulus Technologies AG, Hanauer Landstrasse 107, 63796 Kahl am Main, Germany}

\author{ A. Lamperti}
\affiliation{CNR-IMM, UoS Agrate Brianza, Via Olivetti 2, 20864 Agrate Brianza, Italy}

\author{ N. Wolff}
\affiliation{ Faculty of Engineering, Institute for Material Science, Synthesis and Real Structure, Kiel University, Kaiserstraße 2, 24143 Kiel, Germany }

\author{A. Lotnyk}
\affiliation{Leibniz Institute of Surface Engineering (IOM), Permoserstraße 15, Leipzig 04318, Germany}

\author{J. Langer}

\affiliation{Singulus Technologies AG, Hanauer Landstrasse 107, 63796 Kahl am Main, Germany}

\author{ L. Kienle}
\affiliation{ Faculty of Engineering, Institute for Material Science, Synthesis and Real Structure, Kiel University, Kaiserstraße 2, 24143 Kiel, Germany }

\author{ G. Jakob}
\affiliation{Institute of Physics, Johannes Gutenberg University
Mainz, Staudingerweg 7, 55099 Mainz, Germany}

\author{ B. Borie}
\affiliation{ 
Spin-Ion Technologies, 10 boulevard Thomas Gobert, 91120 Palaiseau, France 
}%

\author{ A. Kehlberger}
 \affiliation{Sensitec GmbH, Walter-Hallstein-Straße 24, 55130 Mainz, Germany}

\author{ D. Ravelosona}
\affiliation{ 
Spin-Ion Technologies, 10 boulevard Thomas Gobert, 91120 Palaiseau, France 
}
\affiliation{ 
C2N, CNRS, Université Paris-Saclay, 10 boulevard Thomas Gobert, 91120 Palaiseau, France}

\author{M. Kläui}
\affiliation{Institute of Physics, Johannes Gutenberg University
Mainz, Staudingerweg 7, 55099 Mainz, Germany}
\date{\today}

\begin{abstract}


This study reports the effects of post-growth He$^+$ irradiation on the magneto-elastic properties of a $Ni$ /$Fe$ multi-layered stack.
The progressive intermixing caused by He$^+$ irradiation at the interfaces of the multilayer allows us to tune the saturation magnetostriction value with increasing He$^+$ fluences, and even to induce a reversal of the sign of the magnetostrictive effect. Additionally, the critical fluence at which the absolute value of the magnetostriction is dramatically reduced is identified. Therefore insensitivity to strain of the magnetic stack is  nearly reached, as required for many applications.  
All the above mentioned effects are attributed to the combination of the negative saturation magnetostriction of sputtered Ni, Fe layers and the positive magnetostriction of the Ni$_{x}$Fe$_{1-x}$ alloy at the intermixed interfaces, whose contribution is gradually increased with irradiation. Importantly the irradiation does not alter the layers polycrystalline structure, confirming that post-growth He$^+$ ion irradiation is an excellent tool to tune the magneto-elastic properties of magnetic samples. A new class of spintronic devices can be envisioned with a material treatment able to arbitrarily change the magnetostriction with ion-induced "magnetic patterning".

\end{abstract}

\maketitle



The magnetoelastic properties of thin films are of major interest for technological use as well as for scientific investigations. The requirements for the magnetoelastic coefficient ($\lambda_s$) strongly depend on the application. Magnetic sensors often need, for example, strain immunity\cite{ota2018flexible}, i.e. zero magnetostriction, to reduce strain cross-sensitivity, while actuators require giant strain effects, achieved in materials such as $TbFe_2$ (terfenol)\cite{garcia2016magnetic}. One way to obtain the optimal value of the magnetostriction for a specific application, is to use the combination of two or more materials with different magnetic and magnetoelastic properties. Multilayer systems have been widely investigated exploiting the combination of different parameters $\lambda_s$ to achieve a target value\cite{nagai1988properties,senda1989magnetic,rengarajan1997effect,jen2005magnetostriction}. In these studies, atomic intermixing at the multilayer interfaces has been identified to severely influence the total magnetostriction and this interface magnetostriction has been exploited to engineer the total magnetoelastic coupling of the multilayer \cite{nagai1988properties,senda1989magnetic}. In ion-sputtered films, where interface mixing naturally occurs, Nagai et al.\cite{nagai1988properties} were able to change the sign of the magnetostriction of a multilayer magnetic stack by changing the relative thickness of the layers. However, a clear limit to this approach is the lack of control of the inter-layer roughness and degree of intermixing. The latter is indeed fixed by the deposition conditions. This imposes limitations to the usability of this method, as the magnetostriction cannot be arbitrarily changed.

An established technique to induce mixing at interfaces is ion irradiation\cite{zhao2019enhancing,fassbender2004tailoring}. Specifically, the use of light ions such as He$^+$ at energies in the range of $10-30$ $keV$ induces short range atomic displacements without generation of surface defects in the material, which instead is more prevailing for heavy atoms\cite{terris1999ion} such as Ar$^+$ or Ga$^+$ . If compared to alternative techniques to promote atomic diffusion, e.g. annealing, the use of ion irradiation confines the intermixing to the magnetic layer boundaries and avoids mixing with the nonmagnetic seed layers (for details see S1 of the supplementary materials). Additionally, annealing is a uniform process while the local nature of irradiation interaction can be applied to the magnetic patterning of multilayer film system.  For these reasons, ion irradiation is an excellent candidate to obtain a desired value of the magnetostriction in a multilayer, by controlling the vertical extension of the intermixed part. Previous work \cite{juraszek2006swift} reported intermixing induced magnetostriction changes using heavy ions and high energies ($700$ $MeV$). However, the use of these type of atoms can be harmful for thin  magnetic layers\cite{cureton2021review}, whose magnetic properties such as saturation magnetization or perpendicular magnetic anisotropy can be easily degraded. Moreover, the presence of cascade collisions in the material and long-range atomic displacements \cite{devolder2013irradiation} makes the precise control of magnetic properties a difficult task.


\begin{figure*}[ht]
\centering\includegraphics[width=15cm]{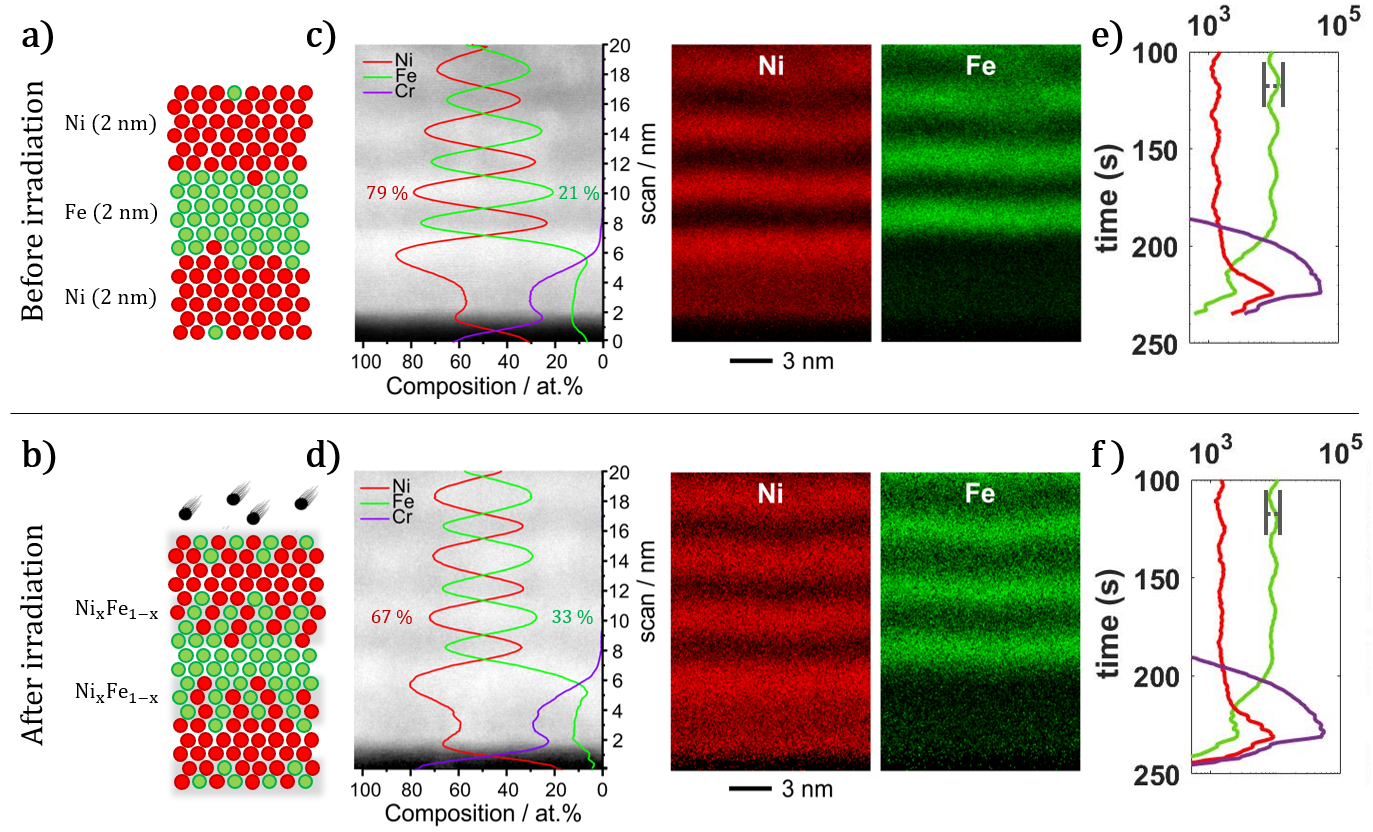}
\caption{\label{fig_1} a) - b) Sketch of the intermixing due to light ion irradiation on a multilayer stack. STEM HAADF micrograph and EDX elemental maps of the Fe/Ni multilayer system before c) and after d) $1\times 10^{16} $ $cm^{-2}$ He$^+$ ion irradiation measured across the first four repetitions on top of the NiFeCr seed layer. The HAADF micrograph is superimposed with a plot of the atomic composition quantified from the EDX measurements.  e) - f) Depth distribution of the elements composing the multilayer structure  obtained with ToF-SIMS in the as-deposited and irradiated samples respectively. }
\end{figure*}

 In this work, we study the effect of progressive intermixing at the interfaces of a $Ni/Fe$ multilayer caused by light-ion irradiation at different fluences. We report that He$^+$ ion irradiation can be used to tune locally the magnetoelastic properties of in-plane magnetized $Ni/Fe$ multilayers, changing the saturation magnetostriction $\lambda_s$ of the magnetic stack from negative to positive. Importantly, we confirm that the above mentioned method not only preserves the layers polycrystalline structure, but also improves the magnetic softness of the material, reducing the coercive field up to $70\%$ and the anisotropy. The key advantages of the proposed method are the high repeatability of the process and the surface uniformity of the magnetic properties. Moreover this technology allows for ion-induced "magnetic patterning",  performing the irradiation through a mask in a similar fashion to semiconductor doping\cite{fassbender2004tailoring,devolder1999patterning}.

The samples have been prepared by magnetron sputtering using a Singulus Rotaris system on a $SiOx/Si$ substrate. A multilayer of $[Ni(2$ $nm)/Fe(2$ $nm)]\times 8$ is sputtered in the presence of a rotating magnetic field of $5$ $mT$ on a $NiFeCr$ $(5$ $nm)$ seed layer and capped with $4$ $nm$ of $Ta$. After that, optical lithography and ion etching have been  used to pattern arrays of circles ($80$ $\mu m$ of diameter and $3$ $\mu m$ of spacing) on the samples in order to probe the local film properties. Multiple copies of the samples have been irradiated at an energy of $20$ $keV$ with different fluences of He$^+$ ions from $5\times 10^{13} $ to $1\times 10^{16} $ $cm^{-2}$. As reported elsewhere for similar irradiation conditions \cite{fassbender2004tailoring}, collision cascades
are absent and the structural modifications are confined to the
vicinity of the ion path in the metal.

At low fluences, it has been shown\cite{devolder2001x} that room temperature irradiation releases strain, whereas, at high fluences, one major structural effect of irradiation is intermixing, as schematically presented in figure \ref{fig_1} a) and b) and confirmed by Montecarlo (TRIM) simulations (figure S3 of the supplementary material).  X-ray diffraction and scanning transmission electron microscopy (STEM) studies indicate in our sample a polycrystalline structure of (110)-textured layers of Fe and (111)-textured layers of Ni  which is not significantly altered by the process of irradiation. In-depth investigation on the structural changes induced by  He$^+$ irradiation and annealing can be found in section S1 of the supplementary material, where atomic diffusion activated by thermal energy is compared with ion irradiation.
According to TRIM simulations\cite{ziegler2010srim}, the majority ($95\%$) of the ions reaches the substrate therefore a uniform intermixing in the vertical direction of the sample is expected, moreover, the effect of ion implantation in the multilayer is negligible.

\begin{figure*}[ht]
\centering\includegraphics[width=16cm]{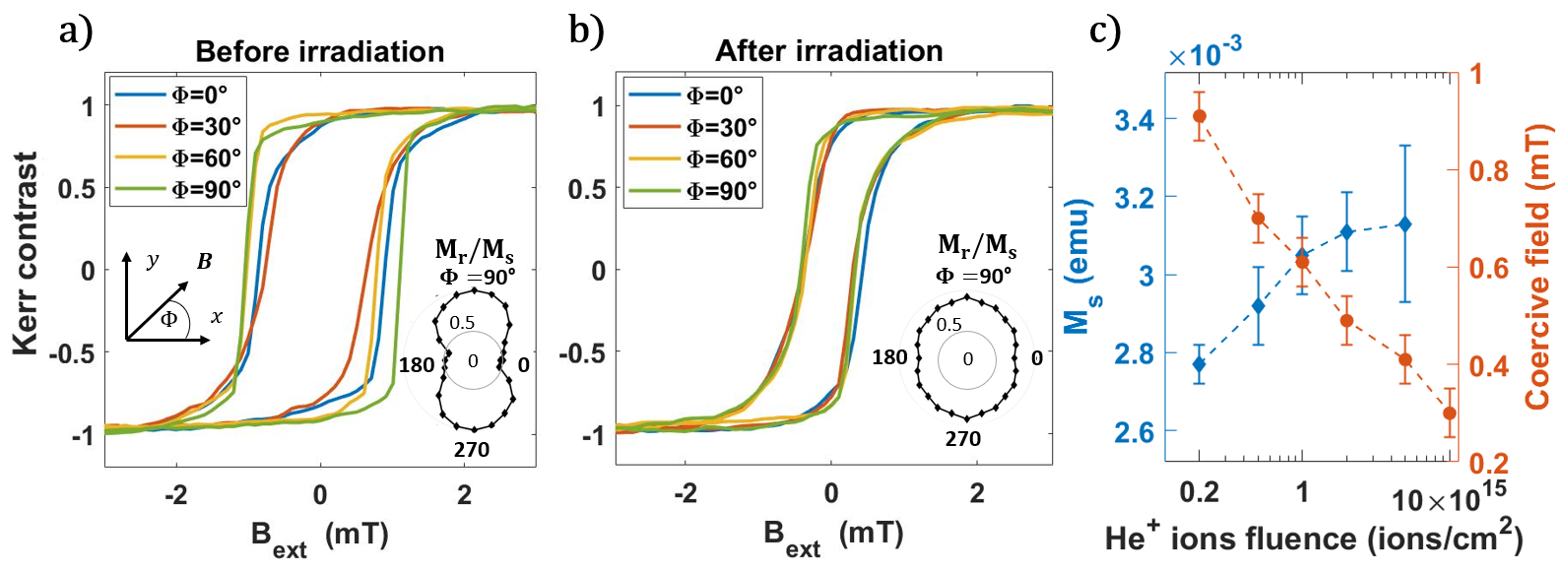}
\caption{\label{fig_03}
a) - b) hysteresis loops as a function of the in-plane magnetic field direction measured by Kerr microscopy, respectively, before and after ion irradiation with a fluence of $1\times10^{16}$ $cm^{-2}$. As inset, angular plot of the remanent magnetization $M_r/M_s$ as a function of the magnetic field angle $\Phi$. c) saturation magnetic moment (light blue) and coercive field (orange) as a function of the fluence of He$^+$ ions during irradiation measured with VSM.}
\end{figure*}

To have a more quantitative estimation of the formation of the alloy for increasing ion fluences, a series of experiments to probe structural and chemical modifications occurring at the layer interfaces caused by ion irradiation were performed and are summarized in figure \ref{fig_1}. Cross sections of Fe/Ni/NiFeCr on SiO$_2$/Si were prepared using the focused-ion-beam method (FIB). High-angle annular darkfield scanning transmission electron microscopy (HAADF-STEM, 80-200 mrad) images were acquired and nanoscale chemical analysis via energy dispersive X-ray spectroscopy (EDX) was performed in STEM mode. A vertical EDX profile across the bottom  layers of the multilayer stack is shown in figure \ref{fig_1} c) and d) together with corresponding EDX maps of the elemental distribution recorded on multilayers before and after He$^+$ irradiation with $1\times 10^{16} $ $cm^{-2}$ fluence, respectively.
After sputtering (figure \ref{fig_1} c) ) the interfaces between the magnetic layers are well defined. The EDX profile of the relative atomic composition indicates 21(2)$\%$ of Fe in a Ni layer  before the irradiation. After irradiation (figure \ref{fig_1} d) ) the ratio of Fe atoms in a Ni layer increases to 33(4)$\%$. This measured stoichiometric change in the layer composition is reflected in the displayed EDX elemental maps by the increased diffuse scattering of signal intensity across the layer interfaces after irradiation. This suggests the formation of an alloy of $Ni_{x}Fe_{1-x}$ at the Ni/Fe interfaces when the different atoms are displaced under the  effect of incoming He$^+$ ions.

Figure \ref{fig_1} e) and f) display the atomic depth  distribution measured by Time-of-Flight Secondary Ion Mass Spectrometry (ToF-SIMS) \cite{benninghoven1987secondary,sodhi2004time, vickerman2013tof, lamperti2013thermal, conte2015role}. The  presence of Fe, Ni and Cr atoms in the multilayer is reported for samples as-deposited and irradiated  with $1\times 10^{16} $ $cm^{-2}$ fluence, respectively. Observing figure \ref{fig_1} e) the position of the periodic oscillations of Ni and Fe appear well defined and have the same periodicity. The peak position, minima of Ni at maxima of Fe, reflects the layer distribution. The atomic distribution after irradiation is shown in figure \ref{fig_1} f).  In this case, the amplitude of Ni and Fe oscillations is significantly attenuated with respect to the as-deposited case. This is again attributed to the intermixing of the atoms in the neighbouring magnetic layers, leading to the formation of $Ni_{x}Fe_{1-x}$ alloy. More details about simulations and  measurements to probe structural modifications can be found in section S1 of the supplementary materials.

The thin film magnetic properties have been measured with Kerr microscopy and Vibrating Sample Magnetometry (VSM). Figures \ref{fig_03} a) and b) show in-plane hysteresis loops before and after the ion irradiation, respectively. In figure \ref{fig_03} a), the as-deposited sample presents different magnetization curves for different angular directions of the magnetic field, indicating the presence of uniaxial crystalline anisotropy $K_u\simeq 100$ $J/m^3$ as can be observed in the inset. The coercivity, measured along the easy axis of magnetization is $0.95(5)$ $mT$. The same magnetic measurements are reported in figure \ref{fig_03} b) for the sample after He$^+$ irradiation of $1\times 10^{16} $ $cm^{-2}$. The magnetic in-plane anisotropy is now negligible, as the the different hysteresis loops overlap. The coercivity is reduced to $0.25(5)$ $mT$. The reduction in coercivity and anisotropy might be related to a possible increase in the concentration of nucleation sites after irradiation, which allow domain formation and switching of the magnetization at lower fields. As the $H_c$ and the magnetic anisotropy are reduced, the magnetic softness of our multilayer is improved by this material treatment.  Figure \ref{fig_03} c) reports systematic measurements of the magnetic properties of our $[Ni(2$ $nm)/Fe(2$ $nm)]\times 8$ multilayer as a function of the He$^+$ fluence during irradiation.  With increasing He$^+$ fluence the magnetic moment of the sample increases by about $15\%$, from 2.8(1) to 3.1(2)$\times10^{-3}$ $emu$. As reported elsewhere\cite{srivastava2006swift}, this is an indication of increased level of intermixing of our magnetic layers (Ni and Fe).

In order to evaluate the potential of ion irradiation to finely tune the magnetoelastic properties of a magnetic multilayer, the effective magnetic anisotropy in our sample has been measured under the application of mechanical strain by three-point bending method as previously reported  \cite{masciocchi2021strain}. Here the substrate is bent to exert a uniaxial strain on the sample. Since the magnetization is coupled to the external strain via the expression of the anisotropy energy\cite{bur2011strain} one way to probe the effect of the strain is to observe changes in the hysteretic behavior before and after mechanical deformation. More details can be found in section S2 of the supplementary material. The  expression of the magnetoelastic anisotropy depends on the saturation magnetostriction $\lambda_s$ of the material according to\cite{finizio2014magnetic}

 \begin{equation} \label{eq_strain_eanis}
K_{ME}=\frac{3}{2}\lambda_s Y \epsilon,
\end{equation}

where $Y$ is the Young's modulus and $\epsilon$ is the uniaxial tensile strain. If the directions of the crystalline and magnetoelastic uniaxial anisotropy are such that $K_u\perp K_{ME}$,  the strain dependent effective anisotropy $K_{eff}$ measured in the system can be written as sum of two terms according to\cite{martin2009local} 

 \begin{equation} \label{eq_energy_tot}
K_{eff}=K_{u}+K_{ME}.
\end{equation}

As the sign of $K_{ME}$ can be negative or positive, depending on the value of $\lambda_s$, the total magnetic anisotropy can, respectively, increase or decrease in the presence of strain. To quantify $K_{eff}$  hysteresis loops are measured using Kerr microscopy, where the magnetic field and the tensile strain are applied along the fixed direction $\Phi=0^\circ$.

\begin{figure}[h!]
\centering\includegraphics[width=8cm]{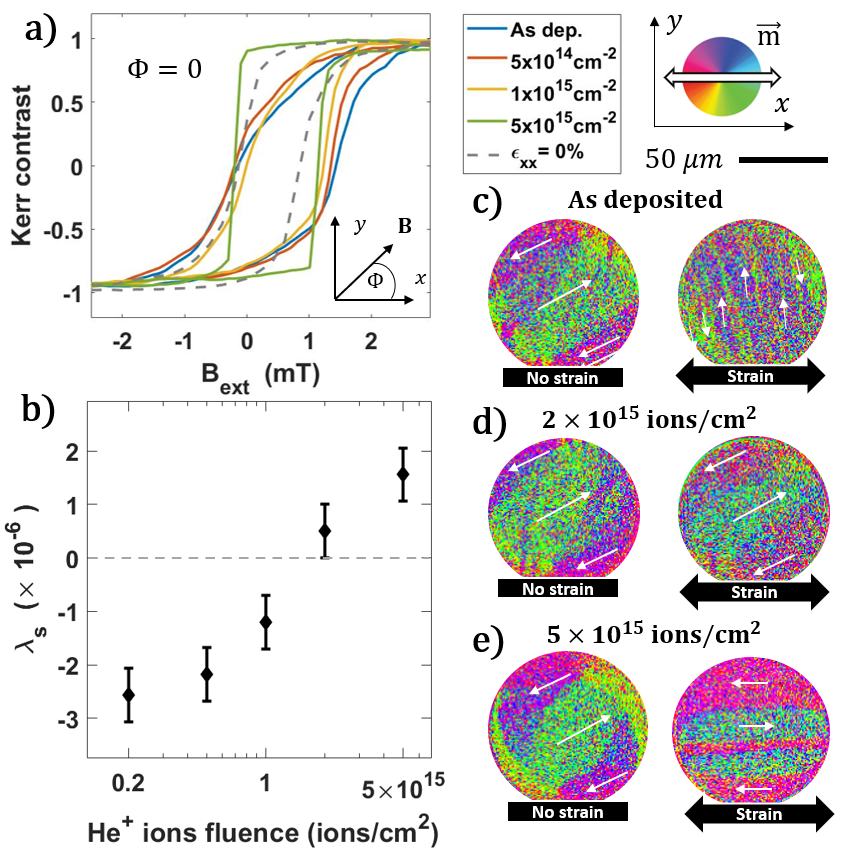}
\caption{\label{fig_02} a) hysteresis loops measured along the direction of the applied strain ($\epsilon_{xx}=0.06\%$) for different fluences of He$^+$ ions (solid lines) are compared with the unstrained magnetic loop (dashed line). b) measured saturation magnetostriction $\lambda_s$ as a function of the ion fluence. Kerr microscope images of the remanent magnetic domain state respectively before (left) and after (right) the application of strain are compared for c) as deposited case, d) intermediate value of irradiation and e) strong value of irradiation. White arrows represent the direction of the in-plane magnetization in the domains according to the color wheel.}
\end{figure}

 The hysteresis loops measured along the direction of the applied  strain $\epsilon_{xx}=0.06\%$ are reported  in figure \ref{fig_02} a) for samples irradiated with different fluences of ions. In response to the applied strain, the irradiated samples have a different magnetic anisotropy field. By comparison with the magnetization curve in the absence of strain (dashed line) two  potential scenarios are identified. When a tensile strain increases the anisotropy field in the direction parallel to $\epsilon_{xx}$, $K_{ME}$ and $\lambda_s$ are negative.  Our sample exhibits negative magnetoelastic coupling in the as-deposited state.  On the other hand, if the strain direction becomes an easy-axis of magnetization (reduced anisotropy field), $K_{ME}$ and $\lambda_s$ are positive. This behavior is reported for larger fluences in the same magnetic stack.

As the difference between magnetic loops before and after the application of strain is proportional to the magnetoelastic anisotropy, the saturation magnetostriction ($\lambda_s$) of our magnetic multilayer can be estimated\cite{choe1999high, raghunathan2009comparison, hill2013whole} using eq. \ref{eq_strain_eanis} and eq. \ref{eq_energy_tot}. Figure \ref{fig_02} b) shows the saturation magnetostriction as a function of the fluence of He$^+$ ions. In agreement with the behavior of the magnetic loops, the value of magnetostriction of the as deposited $Ni/Fe$ multilayer is $-2.6(5)\times 10^{-6}$ . He$^+$ fluences larger than $5\times 10^{14}$ $cm^{-2}$ gradually reduce the absolute value of magnetostriction that then increases through positive values. The change of sign of the magnetoelastic coupling occurs for fluences between $1\times 10^{15}$ and $2\times 10^{15}$ $cm^{-2}$. 


An additional confirmation of the magnetic behavior of the magnetic stack under strain is obtained by imaging domain formation using the magneto-optical Kerr (MOKE) effect. A vector image of the in-plane magnetization is obtained by the sum of horizontal and vertical components of the magnetic contrast. The MOKE images shown in figure \ref{fig_02} c) - e) present how the preferred direction of magnetization changes before and after the application of $0.06\%$ uniaxial strain in $80$ $\mu m$ disk patterned samples. This particular shape has been chosen since it minimizes the in-plane shape anisotropy. The remanent magnetic domain pattern of the multilayer as-deposited is presented in figure \ref{fig_02} c). Before the application of strain (left) the magnetization aligns to the crystalline anisotropy. After the application of strain magnetic domains orient along the $y$ direction, perpendicular to the uniaxial strain $\epsilon_{xx}$. This is a clear experimental proof of the development of stress induced magnetic anisotropy $K_{ME}\simeq-450$ $J/m^3$ that overcomes the initial anisotropy direction. $K_{ME}$ is perpendicular to the tensile strain direction due to the negative sign of the magnetostriction. The domain structure of the sample irradiated with a He$^+$ dose of $2\times 10^{15}$ $cm^{-2}$ is displayed in figure \ref{fig_02} d), where a value of magnetostriction close to zero is measured. In this case, the orientation of the magnetization is almost unchanged by the presence of strain, meaning that $K_{ME}$ is negligible, compared to the crystalline anisotropy of the material $K_u\simeq100$ $J/m^3$. For higher values of fluences as reported in figure \ref{fig_02} e), the effects of strain on the remanent magnetization state become again significant. This time  the dominant magnetic anisotropy contribution in the system is $K_{ME}\simeq 280$ $J/m^3$ as the domains orient along the x direction, parallel to the applied strain $\epsilon_{xx}$. Thus, the magnetoelastic coupling of the stack has been altered using ion irradiation obtaining values of magnetostriction that range from negative to positive.

As previously reported \cite{senda1989magnetic},  the small value of magnetostriction in our (as-deposited) periodic  system is caused by the balance among the negative magnetostriction of Ni ($\lambda_s^{Ni}=-30\times 10^{-6}$) and  Fe ($\lambda_s^{Fe}=-9\times 10^{-6}$) and the strongly positive magnetostriction of $Ni_{x}Fe_{1-x}$ alloy film ($\lambda_s^{NiFe_{50}}=19\times 10^{-6}$) with a relative composition close to $50\%$\cite{bozorth1993ferromagnetism,cullity2011introduction}. As shown by STEM-EDX measurements, a more intermixed interface region of $Ni_{x}Fe_{1-x}$ is formed at the boundary between Ni and Fe layers by He$^+$ ion irradiation. Hence, the thickness of the positive magnetostrictive alloy increases proportionally to the fluence of the He$^+$ ions during irradiation, as also confirmed by ToF-SIMS measurements. This gradually shifts the magnetostriction of the full stack to positive values. A common way to describe the effective magnetostriction in the presence of intermixing is \cite{nagai1988properties, hollingworth2003magnetostriction, favieres2007interface, jen2005magnetostriction, senda1989magnetic,rengarajan1997effect,jen2004anisotropic}
 
 \begin{equation} \label{eq_free_ener}
\lambda_s=\frac{\lambda_s^{Ni}+\lambda_s^{Fe}}{2}+\left(2\lambda_s^{Ni_{x}Fe_{1-x}}-\lambda_s^{Ni}-\lambda_s^{Fe}\right)\frac{t_{Ni_{x}Fe_{1-x}}}{t_p},
\end{equation}
 
 where $t_p=t_{Ni}+t_{Fe}=4$ $nm$ is the period thickness,  $t_{Ni_{x}Fe_{1-x}}$ describes the thickness of the alloy originated by the intermixing and $\lambda_s^{Ni_{x}Fe_{1-x}}$ is the saturation magnetostriction of the intermixed alloy. After deposition in similarly sputtered Ni/Fe multilayers\cite{nagai1988properties}  $t_{Ni_{x}Fe_{1-x}}$ has been estimated to be around $0.9$ $nm$, under the assumption ${t_{Fe}}/{t_p}=0.5$. Using this value of $t_{Ni_{x}Fe_{1-x}}$ eq. \ref{eq_free_ener}  returns $\lambda_s=-3.2(2)\times10^{-6}$, close to the measured value after deposition. Moreover, the amount of induced intermixing caused by He$^+$ ions can be estimated using eq. \ref{eq_free_ener}.  The calculated $t_{Ni_{x}Fe_{1-x}}$ value is $1.13(5)$ $nm$ at the magnetostriction compensation value ($\lambda_s=0$) and $1.22(5)$ $nm$ for the highest fluence, where the magnetostriction is positive due to the dominant effect of the alloy. This corresponds to $25\%$ increase in the alloy thickness induced by He$^+$ between $1\times 10^{15}$ and $2\times 10^{15}$ $cm^{-2}$, in agreement with the information extracted from STEM-EDX and ToF-SIMS measurements.

In conclusion, this manuscript presents an experimental investigation in to the magnetoelastic properties of sputtered $Ni/Fe$ multilayers after He$^+$ ion irradiation. Using different experimental techniques for structural analysis, the presence of moderate roughness and alloying is observed after sputtering at the Ni/Fe interface. This can justify the small negative value of magnetostriction in the as-deposited state.
In the same way, it was found that light ion irradiation promotes the intermixing of the sputtered layers at the interfaces proportional to the ion fluence. This process can explain the reported changes in the saturation magnetostriction of the magnetic stack. The increasing fluence of the irradiating ions progressively changes the saturation magnetostriction inducing a change in sign of the magnetoelastic coupling of the material, from negative to positive for high fluences. Remarkably, strain insensitivity on the magnetic properties of the proposed material can be obtained with ion fluences between $1\times 10^{15}$ and $2\times 10^{15}$ $cm^{-2}$. Importantly, the polycrystalline structure of the layers is confirmed to be unchanged after the used irradiation conditions.

As a result, post growth He$^+$ ion irradiation has been demonstrated to be an excellent tool that allows to fine-tune the magneto-elastic properties of multilayer magnetic samples. Accordingly, this technique can be expected to be the next generation of material treatment offering the possibility to have local patterning of magnetostriction with high control and flexibility, allowing the realization of highly demanding applications.


\begin{acknowledgments}
The authors acknowledge Prof. J. McCord from Kiel University for fruitful discussions. This project has received funding from the European Union’s Horizon 2020 research and innovation program  under  the  Marie  Skłodowska-Curie  grant  agreement  No  860060  “Magnetism  and  the effect of Electric Field” (MagnEFi), the Deutsche Forschungsgemeinschaft (DFG, German Research Foundation) - TRR 173 - 268565370 (project A01 and B02),  the DFG funded collaborative research center (CRC)1261 / project A6  and the Austrian Research Promotion Agency (FFG). The authors acknowledge support by the chip production facilities of Sensitec GmbH (Mainz, DE), where part of this work was carried out and the Max-Planck Graduate Centre with Johannes Gutenberg University.
\end{acknowledgments}

\section*{Author Declarations}
 The following article has been submitted to Applied Physics Letters. After it is published, it will be found at \textit{publishing.aip.org} .
\subsection*{Conflict of interest }
The authors have no conflicts to disclose.

\section*{Data Sharing Policy }
The data that support the findings of this study are available from the corresponding author upon reasonable request.





\nocite{*}
\bibliography{bibliography}

\newpage



\end{document}


\preprint{}

\title[\textbf{Suppl. material} - Control of magnetoelastic coupling in Ni/Fe multilayers using  He$^+$  ion irradiation]{Supplementary material}

\date{\today}

\maketitle

\subsection*{\textbf{S1} - Intermixing characterization and alloy composition}

\begin{figure}[h!]
\centering\includegraphics[width=16cm]{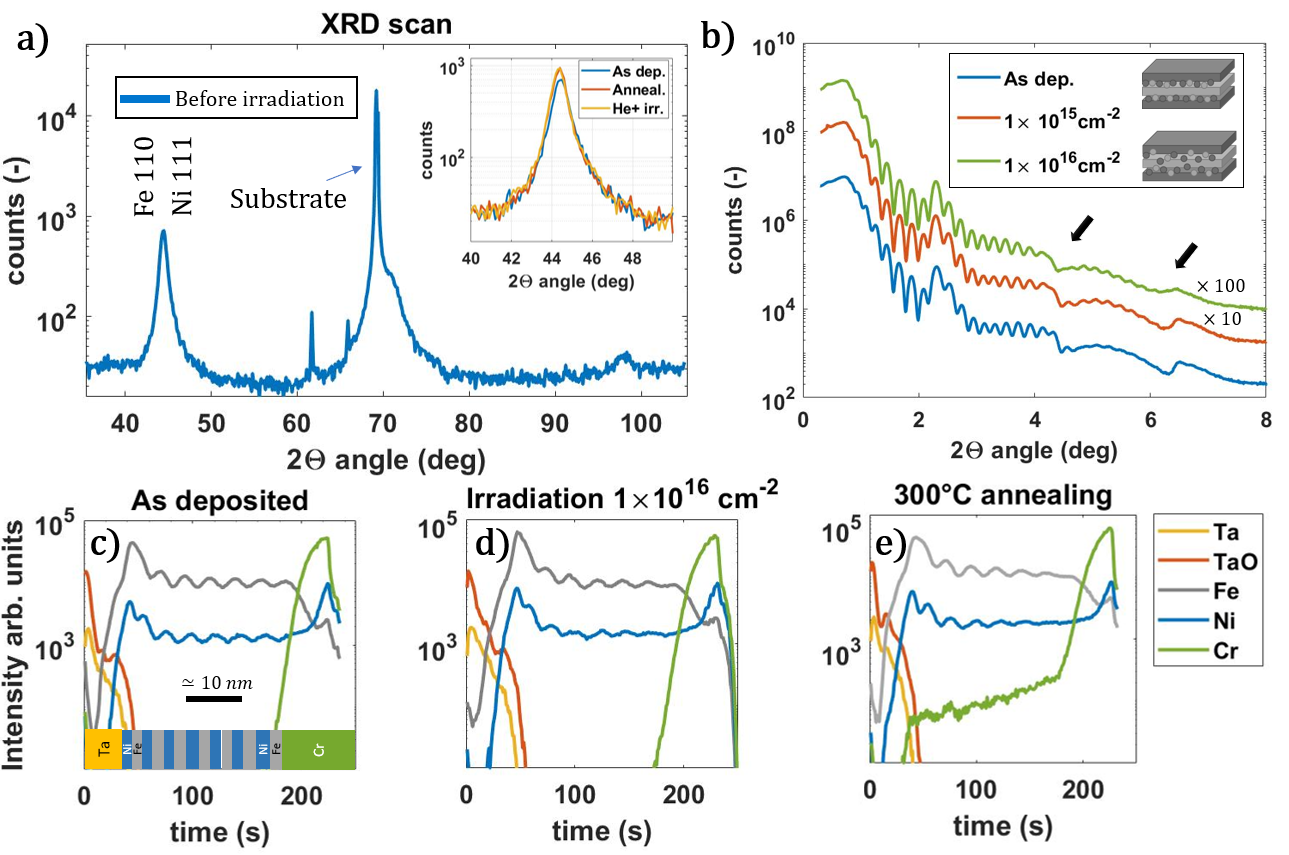}
\caption{\label{fig_S02} a) XRD angular scan of the Ni/Fe multilayer sample after sputtering. In the inset:  Fe 110/Ni 111 peak of the multilayer as-deposited, annealed and after irradiation. b) X-Ray reflectometry (XRR) measurement for a multilayer of $[Ni(2$ $nm)/Fe(2$ $nm)]\times 8$ irradiated with different He$^+$ fluences. The changes in the curves indicate increasing intermixing at the interfaces of our multilayer with increasing ion fluences. c),  d) and e) ToF-SIMS measurements for multilayer as-deposited, after irradiation and thermal annealing respectively. }
\end{figure}

In figure S\ref{fig_S02} a) and b) measurements obtained by X-Ray Diffraction (XRD) are presented, in order to probe the crystalline structure of our multilayer stack. Figure S\ref{fig_S02} a) contains an XRD angular scan of the Ni/Fe multilayer and confirms that our sputtered layers are textured. In figure S\ref{fig_S02} b) an X-Ray reflectivity (XRR) angular scan is shown for as-deposited state and samples with selected irradiation fluences. The best fit considers a relative roughness of the layers of $\simeq1$ $nm$ for the as deposited case. All curves contain two types of periodic oscillation. The short period oscillation with $0.2^{\circ}$ period correspond to the full thickness of the stack ($41$ $nm$ in total).  The long period oscillations around 2, 4 and $6^{\circ}$, correspond to the ML repetitions $t_p=4$ $nm$ (black arrows in figure S\ref{fig_S02} b) ). The amplitude of this oscillations is progressively reduced as the fluence of He$^+$ is increased during irradiation. In figure S\ref{fig_S02} b) this effect of increasing irradiation is best visible for the peak at 6.2$^\circ$. The data suggest a degradation of the Ni/Fe interfaces, indicating an increased level of intermixing. This is qualitatively confirmed by a fitting model where the layer roughness of Ni and Fe is increased. This is in agreement with our ToF-SIMS and STEM measurements. 

A different method to promote atom diffusion in magnetic materials is the use of thermal energy provided by annealing\cite{Annealing}. We compare the effects of  He$^+$ ion irradiation with the annealing in vacuum at 300$^\circ$C for 4.5 hours of our magnetic multilayer.
In the inset of figure S\ref{fig_S02} a) the 110/111 reflection peak is compared for multilayer as-deposited, after annealing and after irradiation with fluence of $1\times10^{16}$ $cm^{-2}$.  The data suggest that the crystalline texture is not altered significantly after these two material treatments.

The atomic diffusion, caused by the adopted material treatment, can be observed in the ToF - SIMS measurements by comparing figure S\ref{fig_S02} c) with figure S\ref{fig_S02} d) - e) after irradiation and the annealing, respectively. As mentioned in the manuscript, the irradiation promotes intermixing as the oscillations in the signals of Fe and Ni are attenuated (fig. S\ref{fig_S02} d) ) with respect to the as deposited case. This effect can be similarly observed after the annealing treatment (fig. S\ref{fig_S02} e) ).  However a clear difference between the figures can be seen in the signal of Cr (from the seed layer). After irradiation the atomic diffusion is confined at the layer interface, instead after annealing the intermixing is long range and involves the non-magnetic NiFeCr seed layer. In addition to that, the coercive field and the magnetic anisotropy of the multilayer are unchanged after the annealing, in contrast to the improved magnetic softness reported after the irradiation. The above mentioned differences between the two material treatment can be attributed to the different activation mechanism for atomic displacement: kinetic energy for irradiation and thermal energy for annealing.

\begin{figure}[h!]
\centering\includegraphics[width=12cm]{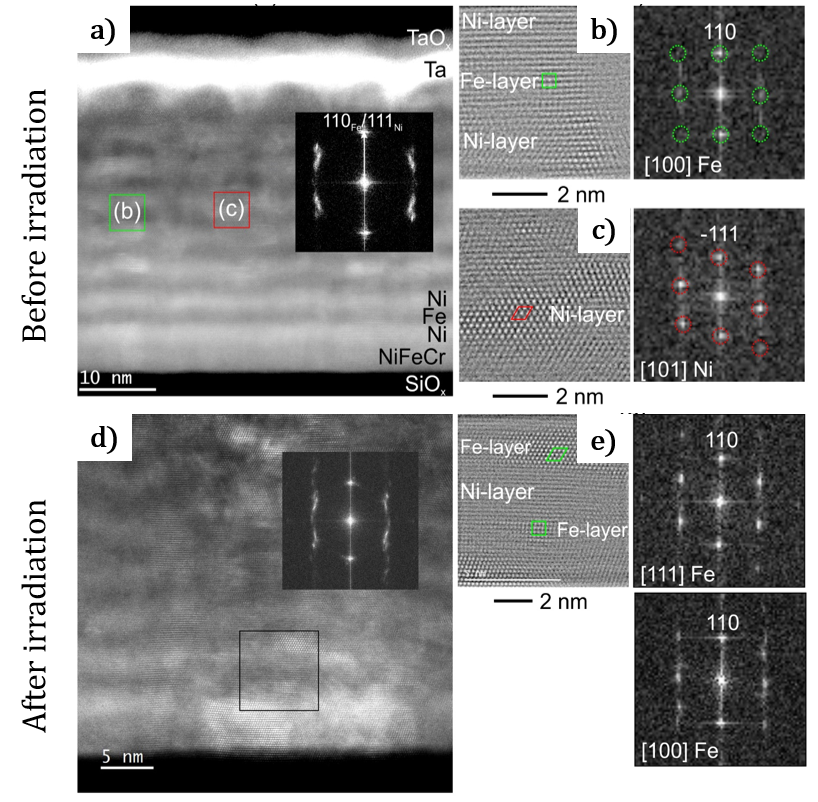}
\caption{\label{fig_S04} High-Resolution STEM micrographs of the Fe/Ni multilayer system before and after He$^+$-ion irradiation. a) repetitions of (110)-textured layers of Fe and (111)-textured layers of Ni are evidenced by specific Z-contrast and individual Fast Fourier Transforms (FFT) of regions b) and c). d) repetitions of Fe and Ni layers showing the identical crystalline texture after irradiation by comparison of FFT images. e) Noise-filtered micrograph displaying the atomic structure of the multilayers. The structural motifs of [100] Fe and [111] Fe are shown for crystalline regions within the Fe-layers.  }
\end{figure}

In addition to ToF-SIMS measurements, the level of intermixing caused by ion irradiation was characterized by Scanning Transmission Electron Microscopy (STEM). Cross sections of Fe/Ni/NiFeCr on SiO$_2$/Si were prepared using the focused-ion-beam method (FIB). STEM images were acquired in high-angle annular dark field (HAADF) mode on a probe CS(spherical aberration coefficient)-corrected Titan$^3$ G$^2$ 60–300 microscope operating at an accelerating voltage of 300 $kV$ using a probe-forming aperture of 25 $mrad$ and annular ranges of 80-200 mrad on the detector. Nanoscale chemical analysis via energy dispersive X-ray spectroscopy (EDX) was performed in STEM mode using a Super-X detector setup with 4 symmetrically aligned in-column detectors.
The  structure of our multilayer is polycrystalline with (110)-textured layers of Fe and (111)-textured layers of Ni as can be seen in figure S\ref{fig_S04}. The structural motif of [100] Fe with (110) out-of-plane orientation was evidenced by Fast Fourier Transforms (FFT)  on a crystalline region within a Fe-layer. Within the Ni-layers, the dominant structural motif of [101] Ni was observed with (111) out-of-plane orientation before He$^+$-ion irradiation as shown in figure S\ref{fig_S04} a) - c). The same measurements have been repeated after ion irradiation with a fluence of $1\times10^{16}$ $cm^{-2}$. In this case the polycrystalline multilayers of Fe and Ni after irradiation  ( FFT images in figure S\ref{fig_S04} d) - e) ) show the identical crystalline texture compared to the as-deposited state  (figure S\ref{fig_S04} b) - c)  ), allowing to exclude significant changes to the crystalline structure after the used irradiation treatment.

\begin{figure}[h!]
\centering\includegraphics[width=16cm]{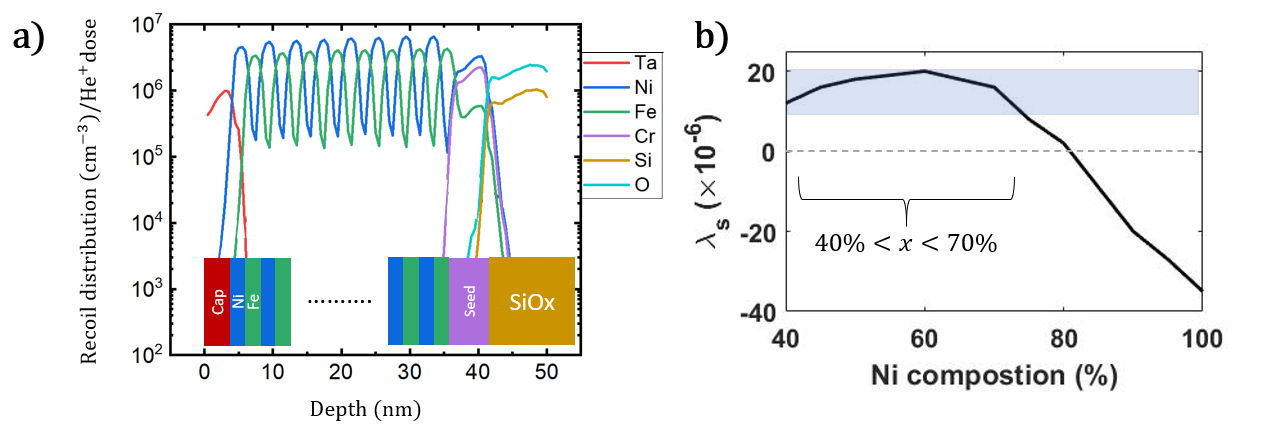}
\caption{\label{fig_S03} a) Montecarlo simulation using the software TRIM\cite{ziegler}. The atomic distribution of different elements after the collision with incoming ions is shown along the vertical depth of the multilayer. The results are  normalized by the incoming fluence of ions. b) saturation magnetostriction $\lambda_s$ of a $NiFe$ alloy as a function of the Ni composition ($\%$). Data points from Judy et al.\cite{fabrication}.  }
\end{figure}

  Montecarlo (TRIM) simulations were performed. Using TRIM simulations it is possible to calculate kinetic phenomena associated with the ion’s energy loss: in our case target atom displacement (normalized by the incoming ion fluence) as a function of the vertical depth of the sample. The system is initialized with perfect interfaces and the kinetic energy of the incoming ions is set to 20 $keV$. The results of TRIM simulations are presented in figure S\ref{fig_S03} a) . The solid lines represent the recoil atomic distribution after the collision with He$^+$ ions. In the overlapping region of two curves we have coexistence of different atomic species (intermixing/alloying). The simulations suggests that the displacement is uniform through the magnetic stack for the selected ion energy, we can therefore expect the same amount of intermixing at each Ni/Fe interface. 
 Furthermore we do not see any significant intermixing of the non-magnetic capping and seed layers with the magnetic stack. This is most likely an effect of the short-range nature of the collisions with He$^+$ ions \cite{Fassbender}. The outcome of the simulations is in line with the ToF-SIMS and STEM-EDX measurements.
 
 In the manuscript we attribute the changes in the magnetoelastic coupling to the intermixing induced by ion irradiation. To describe the magnetostriction of a multilayer system in the presence of intermixing we can use the expression\cite{model}

 \begin{equation} \label{eq_free_ener}
\lambda_s=\frac{\lambda_s^{Ni}+\lambda_s^{Fe}}{2}+\left(2\lambda_s^{Ni_{x}Fe_{1-x}}-\lambda_s^{Ni}-\lambda_s^{Fe}\right)\frac{t_{Ni_{x}Fe_{1-x}}}{t_p},
\end{equation}

where we ascribe the changes of magnetostriction after the irradiation process to the combination of $\lambda_s$ of three different materials: Ni , Fe (which are the sputtered layers) and the $Ni_{x}Fe_{1-x}$ alloy at the interfaces (induced by ion irradiation). In our case the thickness $t_{Ni_{x}Fe_{1-x}}$ grows under the effect of irradiating ions. The predicted values by equation \ref{eq_free_ener}  are obtained considering, for the formed alloy, the value of magnetostriction for relative composition $x=50\%$. All the values are reported in table \ref{tab_material_film}. In realistic conditions, the amount of intermixing will be changing gradually at the interface. Consequently $\lambda_s^{Ni_{x}Fe_{1-x}}= \lambda_s^{Ni_{x}Fe_{1-x}} (x)$ will not be constant. 

The approximation used is justified by the content of figure S\ref{fig_S03}. Montecarlo simulations indicate that the transition between the atomic distribution of the sputtered materials (Ni and Fe) is exponential as shown in figure S\ref{fig_S03} a). This indicates that the formed alloy is confined at the interfaces. Additionally the magnetostriction of permalloy with relative Ni composition between $x=40-70\%$ does not deviate significantly from the value in table \ref{tab_material_film}. This is because the magnetostriction of permalloy has a local maximum around $50\%$ relative composition as can be seen in figure S\ref{fig_S03} b). Therefore, a constant value of $\lambda_s^{Ni_{x}Fe_{1-x}}$ is expected to give consistent results, as has been the case for previous works\cite{model,nagai1988properties}.

Discrepancies between the calculated values and the measured ones, can be attributed to surface and interface effects which sum to the presence of the intermixed layer.

\subsection*{ \textbf{S2} - Evaluation of Magnetostriction}

\begin{figure}[h!]
\centering\includegraphics[width=14cm]{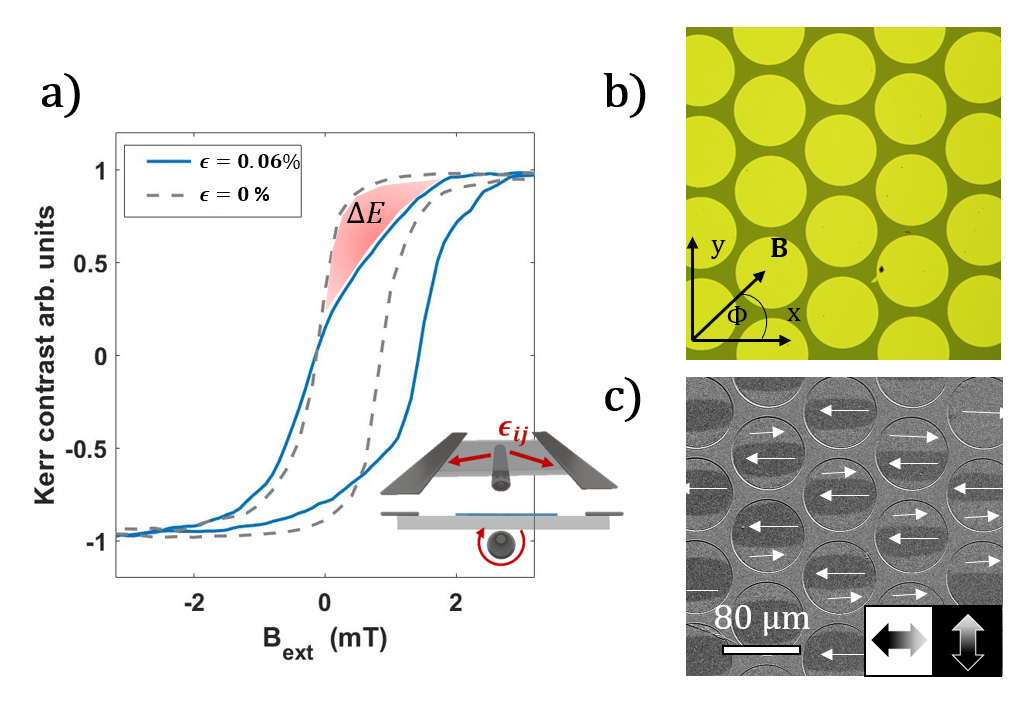}
\caption{\label{fig_S01} a) hysteresis loops measured with Kerr microscopy that are used to estimate $\lambda_s$. Inset: schematic of the sample holder used to strain the substrate. b) optical microscope image of the sample patterned in array of $80$ $\mu m$ diameter and c) magnetic domains imaged with longitudinal configuration of polarized light.  }
\end{figure}

 All layers are deposited by magnetron sputtering ( using a Singulus Rotaris system). The substrate is $1.5$ $\mu m$ SiOx on top of $625$ $\mu m$ undoped Si. The magnetic material was sputtered in a rotating magnetic field of $50$ $Oe$. Our film are structured using optical lithography and Ar$^+$ ion etching into a circular pattern of $80$ $\mu m$ diameter and $3$ $\mu m$ of spacing as shown in figure S\ref{fig_S01} b). This design has been chosen to probe the local magnetic properties of the film while, at the same time, minimizing the shape anisotropy contribution.
 
 The magnetic measurements were performed using Kerr microscopy with a $20\times$ objective and a white light source. Coils for in-plane magnetic field are used. We measure the hysteresis loops detecting differential contrast changes in the magneto-optical Kerr effect (MOKE) in a longitudinal configuration of the polarized light. Both longitudinal and transversal configuration are instead used to image the magnetization state (domains) in a grey scale sum image as can be seen in figure S\ref{fig_S01} c).

To obtain information about the magnetoelastic properties of the material, the substrate was bent mechanically with a 3 point bending sample holder, as shown schematically in the inset of figure S\ref{fig_S01} a). A square sample of 1 by 1 cm is vertically constrained on two sides and pushed uniformly from below by a cylinder that has an off-centered rotation axis. The device generates a tensile strain in the plane of the sample up to $0.1$ $\%$ when the cylinder is rotated by 90$^\circ$. The strain is mostly uniaxial and has been measured with a strain gauge on the substrate surface. 

Magnetic hysteresis loops are recorded before and after the application of the tensile strain and are used to estimate the saturation magnetostriction of the material. As  previously reported\cite{spinvalve,paperarea} the magnetic anisotropy $K_{u}$ is linked to the energy stored in the magnetization curves. For example the (uniaxial) magnetic anisotropy energy is given by the area enclosed between the magnetic loops measured along two in-plane directions perpendicular to each other.  If then the strain in the film is non-zero, the magneto-elastic coupling contributes in principle to the effective anisotropy. Two hysteresis loops measurements, before and after the application of strain, are sufficient to estimate $\lambda_s$. Indeed the total anisotropy of the system is $K_{eff}=K_{u}$ and $K_{eff}=K_{u}+K_{ME}$ before and after the application of strain, respectively. The magnetoelastic anisotropy $K_{ME}=\frac{3}{2}\lambda_s Y \epsilon$ is linked to reversible part of the hysteresis loops (close to the saturation) according to

 \begin{equation} \label{eq_strain_eanis}
K_{ME}=M_s \Delta E=\frac{3}{2}\lambda_s Y \epsilon
\end{equation}

where $\Delta E$ is the anisotropy energy measured by the difference in area below the strained and unstrained curves. This corresponds to the reversible part, i.e. the red marked area in figure S\ref{fig_S01} a). The experimental values of magnetostriction were calculated using the values of the the Young's modulus ($Y$) and saturation magnetization ($M_s$) of the stack taken from literature and reported in table \ref{tab_material_film}.

\begin{table}[h!]
    \centering
    \begin{tabular}{||c c c c||} 
 \hline
    Material & $M_{s}$ $ (T)$  & $\lambda_s$ x$10^{-6}$ & $Y$  ($GPa$) \\ [0.5ex] 
 \hline\hline
 $Fe$ & 2.15 &	-9 & 211\\
  \hline

 $Ni $ &  0.55 &	-30 & 180 \\ 
  \hline
  $Ni_{50} Fe_{50}$ & 1.5 &	19 & 200 \\ 

 \hline
\end{tabular}
\caption{Parameters from literature\cite{nagai1988properties,cullity2011introduction,bur2011strain,bozorth1993ferromagnetism,klokholm1981saturation} of the magnetic materials after deposition (no irradiation). Here, $M_s$ is the saturation magnetization, $\lambda_s$ is the saturation magnetostriction and $Y$ is the Young's modulus. The same $Y$ is considered for as-deposited and irradiated samples. }
\label{tab_material_film}
\end{table}

\newpage
